# Rational design and dynamics of self-propelled colloidal bead chains: from rotators to flagella

Hanumantha Rao Vutukuri[a,*,‖], Bram Bet[b], René van Roij[b], Marjolein Dijkstra[c,*],

Wilhelm T. S. Huck[a,*]

[a]Institute for Molecules and Materials, [a]Radboud University, Heyendaalseweg 135, 6525 AJ   Nijmegen, The Netherlands

[b]Institute for Theoretical Physics, Center for Extreme Matter and Emergent Phenomena, Utrecht University, Princetonplein 5, 3584 CC Utrecht, The Netherlands

[c]Soft condensed Matter, Debye Institute for Nanomaterials Science, Utrecht University, Princentonplein 1, 3584 CC Utrecht, The Netherlands

[‖] Present address: Soft Materials, Department of Materials, ETH Zurich, 8093 Zurich Switzerland

Email: hanumantharaov@gmail.com, m.dijkstra@uu.nl, & W.Huck@science.ru.nl

**The quest for designing new self-propelled colloids is fuelled by the demand for simple experimental models to study the collective behaviour of their more complex natural counterparts. Most synthetic self-propelled particles move by converting the input energy into translational motion. In this work we address the question if simple self-propelled spheres can assemble into more complex structures that exhibit rotational motion, possibly coupled with translational motion as in flagella. We exploit a combination of induced dipolar interactions and a bonding step to create permanent linear bead chains, composed of self-propelled Janus spheres, with a well-controlled internal structure. Next, we study how flexibility between individual swimmers in a chain can affect its swimming behaviour. Permanent rigid chains showed only active rotational or spinning motion, whereas longer semi-flexible chains showed both translational and rotational motion resembling flagella like-motion, in the presence of the fuel. Moreover, we are able to reproduce our experimental results using numerical calculations with a minimal model, which includes full hydrodynamic interactions with the fluid. Our method is general and opens a new way to design novel self-propelled colloids with complex swimming behaviours, using different complex starting building blocks in combination with the flexibility between them.**

Active or self-propelled colloidal-particle systems are currently a subject of great interest in soft condensed matter science, owing to their ability to mimic the collective behaviour of complex living systems, but also to serve as model systems to study intrinsically out-of-equilibrium systems.[1-6] Self-propelled particles can exhibit rich collective behaviour, such as clustering, segregation, and anomalous density fluctuations, by consuming internal energy or extracting energy from their local environment in order to generate their own motion.[1-5,7] In recent years, several experimental self-propelled particle systems have been developed, namely Janus spherical particles,[2,8-11] Janus rods,[12] bimetallic rods,[13] granular rods,[14] gear-shaped,[15,16] hetero-dimers,[17] and L-shaped particles,[18] based on different self-propulsion mechanisms. However, most of the experimental studies have been reported on Janus spherical[19] and rod-like particle systems[12]. Moreover, the majority of self-propelled particles move actively by converting the input energy directly into translational motion on a single particle level. As a result, the self-propelled particles exhibit ballistic behaviour at time scales much smaller than the rotational diffusion time and diffusive behaviour at longer time scales, with an effective diffusion coefficient that is many times larger than the equilibrium diffusion coefficient. In this work we ask ourselves whether, and if so how, self-propelled Janus spheres can assemble into bead chains that exhibit rotational motion, possibly coupled with translational motion. Some biological systems display active rotational motion, e.g., certain bacteria confined to two dimensions,[20] and dancing algae.[21] To the best of our knowledge, only a few synthetic systems show active rotation, e.g., a granular gear-shaped particle in an active bacterial bath,[15,16] multi-layered bimetallic rods,[22,23] L-shaped particles,[24] and random clusters of active Janus particles.[25,26] However, the swimming behaviour of these systems cannot be tuned by controlling the internal structure. Here, we develop a new type of internally driven colloidal-particle system, namely self-propelled colloidal bead chains with tunable stiffness. This system is suitable to study dynamics on a single particle level in concentrated systems, and exploits directed self-assembly of simple self-propelled building blocks. Moreover, we show that introducing (semi-)flexibility between the beads in a chain is sufficient to break the time-reversibility constraint to realize flagellum-like or helical-like motion that is powered by the fuel. To the best of our knowledge, flagellum-like motion has not been reported before in internally powered (self-propelled) colloidal swimmers. Only externally powered propellers such as helix-shaped[27-30] particles using rotating magnetic fields, and DNA-linked assemblies of magnetic particles tethered to a red blood cell[31] in biaxial magnetic fields have shown this motion. However, these systems are not suitable for collective behaviour studies because the long-ranged magnetic interactions between individual units are difficult to minimize. Additionally, they do not satisfy the force-free and torque-free condition[28] which is an essential condition for studying internally powered colloidal systems. Moreover, these systems are difficult to fabricate.

Several simulations and theoretical studies have predicted that particle shape and swimming direction can affect the macroscopic behaviour of self-propelled particle systems,[1,2,6,32] but the experimental realization of shape-anisotropic particles in combination with varying propulsion directions scarcely exists. Several synthesis routes have been reported for synthesizing 'passive' colloidal molecules, complex-shaped particles,[33-38] chains of particles using different starting building blocks, e.g., isotropic spherical particles using various linking mechanisms,[34,39-42] Janus particles,[43] and flattened particles.[44] However, making these model systems with self-propelling capabilities remains a challenge. Here, we take an inspiration from the design principles of synthesizing passive colloidal molecules, where the colloidal spherical particles are treated as atoms and the interactions between them are tuned such that they self-assemble into well-defined complex clusters.[33,34,36] We show that self-assembly of the Janus particles into polymer-like chains with tunable stiffness provides a powerful route towards the fabrication of complex self-propelled particle systems.

**Fabrication of permanent bead chains**

We first present our method to create catalytically powered colloidal rotators or active rigid linear bead chains by connecting the individual self-propelled spherical swimmers in such a way that neighbouring swimmers in the chain are always propelling in opposite directions. Our fabrication method consisted of two steps (Fig. 1a): (i) aligning Janus particles into linear chains using a high frequency external AC electric field,[45] and (ii) making these structures permanent using a combination of van der Waals attractions and linkers (i.e., inter-diffusion of polymer chains that are made of particles, and polyelectrolytes) to ensure that the chains remained intact as a chain even after removing the field. Suspensions consisting of 0.02 *vol%* half-side Pt coated polystyrene (PS) particles in deionized water were introduced to a rectangular electric cell (0.1 X 1.0 mm$^2$). Upon application of a high frequency external AC electric field ($E_{rms}$ = 0.01 Vμm$^{-1}$, $f$ = 800 kHz where $E_{rms}$ is the root-mean-square electric-field strength and $f$ is the frequency), the particles assembled into staggered or zig-zag linear chains in the direction of the applied field (SFig. 1b). The difference in the polarizability of both sides (half-dielectric and half-metallic) of the Janus particles caused the particles to align into staggered chains as shown in SFig. 1b. At low particle concentrations and at low field strengths, the stable structure is a string fluid phase that consists of staggered chains of particles in the field direction and a liquid-like order in the perpendicular to the field direction (SFig. 1b).[45] However, these linear structures rely on the presence of external electric fields, i.e., the chains lost their identity (SFig. 1c) when the field was switched off.[37,45] To circumvent this, we used a combination of a strong electric field and a heating step. At high field strengths ($E_{rms}$ = 0.04-0.05 Vμm$^{-1}$, $f$ = 800 kHz), the induced dipolar attractions between the particles are strong enough to push the particles together into distances where the van der Waals attractions take over and bind the particles irreversibly. In particular, the van der Waals attractions between metallic halves of the particles are strong in comparison to those between the polymeric halves of the particles.[46] Due to some irregularities in the Pt coating a small fraction of the polymeric side of a particle can also touch the polymeric side of neighbouring particles. These polymer-polymer connections are the key to making permanent bonds between neighbouring Janus spheres through a heating step that was developed in an earlier study by Vutukuri *et al.*[34,39] In the heating step, the sample cell was heated to 60-65 °C, which is well below the glass transition temperature[47] ($T_g$ ≈ 107 °C) of polystyrene as shown in SFig.2, for about 2-3 minutes using a hot-air stream that was much wider than the sample cell.[34,39] After 2-3 minutes, the field was slowly switched off and we found that the particles were permanently attached to each other in a staggered fashion and acted as a single rigid body. SFig. 3 shows the length distribution of the resulting permanent bead chains. It is important to mention here that the length of the permanent chains can be controlled by varying the distance between the two electrodes as has been developed in our previous study by Vutukuri *et al.*[34]. This procedure could thus significantly increase the monodispersity in length, if necessary, but was not used in the present study.

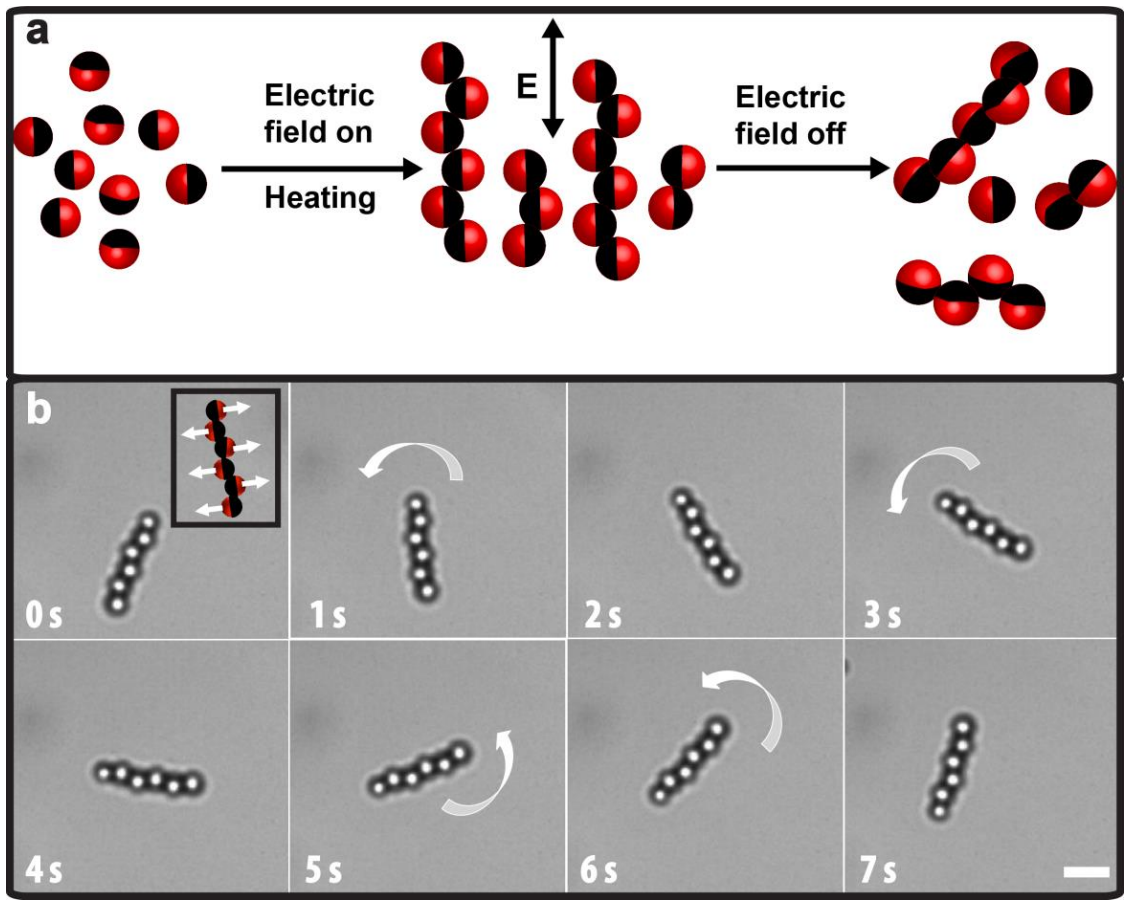

**Figure 1.** (a) Fabrication of active rotators. Schematic diagram illustrating the steps involved in the preparation of active bead chains. Dark side represents the Pt-coated side of the particles. (b) Time evolution of an active rigid colloidal polymer chain or self-propelled rotator in 1.0 vol% $H_2O_2$ solution. Time-lapse optical microscopy images show the circular motion of a bead chain in the presence of the fuel. The arrow depicts the direction of the rotational motion. (Inset) The direction of the propulsion forces acting along the length of the chain. Dark parts represent metal-coated (Pt) side of the particle. Scale bars are 2.0 μm.

**Results and Discussion**

**Dynamics of self-propelled rotators or active rigid bead chains**

After creating the permanent structures, we studied their dynamic behaviour in the presence of $H_2O_2$. We note that it has been reported that in the case of individual half Pt-coated Janus spheres, the platinum-catalyzed decomposition reaction of $H_2O_2$ generates a concentration gradient of ions across the surface of the particles, inducing self-propulsion in the direction of the non-coated surface.[8-10,48] Although the underlying mechanism responsible for the propulsion is still under debate,[49] the dominant propulsion mechanism is probably a combination of diffusiophoresis and self-electrophoresis. [8-10,48] When we transferred the permanent staggered chains into a 1.0 *vol%* $H_2O_2$ solution, the bead chains showed autonomous sustained rotating behaviour on the surface of the bottom wall (Fig. 1b). The rotating movement can be attributed to the fact that the propulsion force on each Janus particle is acting in opposite directions in alternative fashion along the length of the chain (see the inset of Fig. 1b), resulting in a constant torque acting on the chain, which induces a rotational motion of the chain. The time-lapsed bright field images show the rotational motion of a 6-bead chain in 1.0 *vol%* of $H_2O_2$ solution as shown in Fig. 1b (see supplementary Movie S1). The inset of Fig. 2b shows the typical trajectory of the centre-of-mass of the bead chain. In order to characterize the trajectories, we first extracted the centroids of each bead (Janus spheres) in a chain using a particle-tracking algorithm,[50] and then calculated the orientation angle based on the coordinates of the first and the last bead as a function of time (see Materials and Methods section). From this, we calculated the mean-squared angular displacement (MSAD), and the translational mean-squared displacement (MSD) of the centre of mass of the chain as shown in Fig. 2. Due to the persistent angular propulsion, the bead chains are actively rotating, as shown by the cumulative rotational angle $\theta(t)$ continuously increasing with time (inset of the Fig. 2a). We obtained the average angular velocity ($\omega$) and the passive (equilibrium) rotational diffusion coefficient ($D_{r,eq}$) of the rotator by fitting the MSAD curves with the equation $\langle \Delta \theta^2 \rangle = \langle [\theta(t) - \theta(0)]^2 \rangle = \omega^2 t^2 + 2 D_r t$ (see more details in Materials and Methods section). The rotator or active bead chain was propelling with an angular velocity $\omega$ = 0.48 ± 0.03 rad/s. The obtained equilibrium rotational constant,

$D_{r,\ eq}$ = 0. 005 ± 0.004 $rad^2/s$, is consistent with the theoretical value[51] as well as with our numerical calculations (see the bead-shell minimal model in Materials and Methods section). Note that at the same fuel concentration (1.0 *vol%* of $H_2O_2$ solution), singlets propelled with a velocity of 1.58 μm/s (see SFig. 4). In contrast to single self-propelled spheres, our rotators show distinctive MSAD and MSDs. The main distinctive feature of the MSD is an oscillatory behaviour indicating that the rotator performs a spiral-like "spira mirabilis" motion due to thermal fluctuations. As shown in Fig. 2b, the experimental MSD is in a good quantitative agreement with the theoretical predictions as have been reported for a thin rod with not only a self-propulsion force, but also a self-propulsion torque, i.e., a circular swimmer.[52] We note that the angular velocity obtained from the MSAD fitting ($\omega$ = 0.48 ± 0.03 rad/s) is in quantitative agreement with the value obtained from the MSD fitting ($\omega$ = 0.42 ± 0.09 rad/s). Next, we quantified the dynamics of bead chains in terms of angular velocity for different chain length and at different fuel concentrations. For each fuel concentration, we analysed 6 trajectories for each chain length in two separately prepared samples. In Fig. 3, we plot the average angular velocity as a function of chain length, for two different fuel concentrations. It is apparent from Fig. 3 that the chains with an even number of beads exhibit rotations that are much more pronounced than those of chains with an odd number of beads. This can be understood from the forces on the beads: in a perfectly symmetric chain where the propulsion forces are antiparallel, a chain with an even number of beads will experience a net torque, but no net force. In contrast, a chain with an odd number of particles will experience no net torque but will experience a net force. In practice, small asymmetries result in both forces and torques on both types of chains, but the preference of odd-numbered chains for translation and even-numbered chains for rotation is preserved. Both clockwise (CW) and counterclockwise (CCW) rotations are observed for different chain lengths. For longer bead chains (> 3σ, where σ is the particle diameter) the gravitational height of the passive bead chains is $l_g$ >1.0 μm, which is comparable to the particle diameter. As a result, they are confined to a plane where they, once started rotating in a particular direction, tend to continue in that direction. We did not observe any switching between the directions from the CCW to the CW directions and vice versa. On the other hand, for dimer and trimer chains we did observe switching between the rotation directions. This can be attributed to the fact that the gravitational height for shorter bead chains is less than one particle diameter. For instance, the gravitational height of a dimer bead chain is ~ 0.3 σ. As a consequence, shorter chains do not only show small fluctuations out of the 2D plane, but show also full rotations in 3D, thereby switching between CCW and CW rotation directions. For an even number of beads in a chain, the angular velocity decreases with increasing length at constant fuel concentration, as shown in Fig. 3. In order to capture the flow fields generated by the rotators in the fluid, we added uncoated PS spheres of diameter 2.1 μm as tracer particles in the sample cell. We subsequently followed the dynamics of the tracer particles as a function of time by means of a bright field microscope. The trajectories of the tracer particles reveal that the tracer particles in close proximity of the 4-bead rotator show net motion (Fig. 4), while the particles far from the rotator show Brownian motion and the effect of the rotator is negligible in this case. Based on our experimental observations we believe that there is no bulk convective flow present in the system.

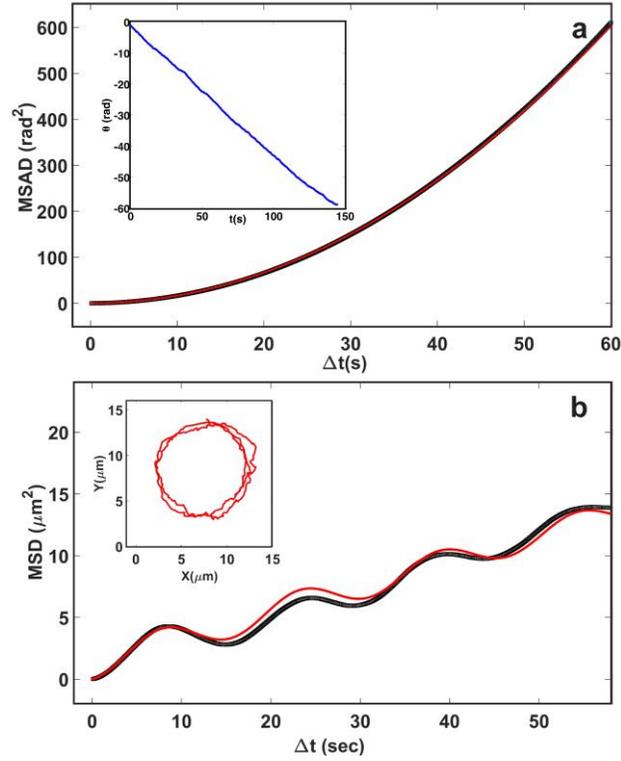

**Figure 2.** Self-propelled 6-bead chain in a 1.0 vol% $H_2O_2$ solution. (a) Mean-squared angular displacement (MSAD) of rigid active bead chain. The black squares represent the experimental measurement, and the solid red curve shows the quadratic fit from the MSAD equation. Top-left inset shows the cumulative rotational angle with time. (b) The typical mean-squared displacement (MSD) of a circular swimmer. The solid red curve is the fit from the MSD equation (eq.5). Top-left inset illustrates the trajectory of the center-of-mass of the bead chain.

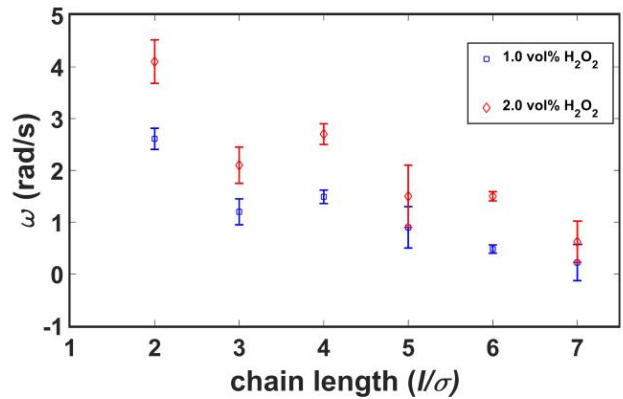

**Figure 3.** Angular velocities ($\omega$) for varying length of active rigid chains and for two different $H_2O_2$ concentrations. Blue squares represent the angular velocity at a fuel concentration of 1.0 vol%, and red diamonds denotes the angular velocity at a fuel concentration of 2.0 vol%.

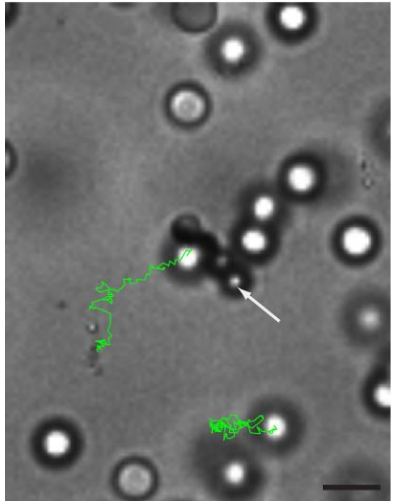

**Figure 4.** The fluid flow generated by the 4-bead rotator in 1.0 vol% $H_2O_2$ solution. Green line represents the trajectory of the PS tracer particles of 2.1 μm size overlaid on a bright field image. The white arrow depicts the rigid 4-bead rotator. The scale bar is 5.0 μm.

**Self-propelled or active semiflexible bead chains**

Next, we study how flexibility between the beads in a chain affects the swimming behaviour in the presence of a fuel. Here we used sterically stabilized PS (polystyrene) Janus particles with a high molecular weight polyvinylpyrrolidone (PVP, $M_w$ = 360 kg mol$^{-1}$) as starting building blocks and repeated the same protocol as was mentioned for the fabrication of active rigid chains. The particles that were sterically stabilized with a higher molecular weight PVP form chains with a semiflexible character[34] as the stabilizing polymers (PVP) that were chemically connected and/or entangled with the PS polymer chains, acted as linkers between the beads. Using Fourier mode analysis, we quantified the flexibility of the resulting chains by estimating the persistence length $l_p$, yielding $l_p \approx 20\sigma$, where σ is the diameter of a bead.[34] When we introduced flexibility between the beads in a chain, the resulting chains showed a completely different swimming behavior consisting of a coupled translational and rotational motion, which strongly resembles flagellum-like motion or helical motion (see supplementary Movie S2). Here, our intention is to refer to the observed swimming behavior as a flagella-like motion, which means that the chain is undergoing an intricate coupling between rotational and translational motion. We note that the rigid bead chains showed only active rotational motion. The time-lapsed bright field images show the sequential shape and orientation changes of an active semiflexible chain, indicated by the change in intensity of beads in the chain as shown in the XY projection in Fig. 5a. We further confirmed this by measuring the intensity profile of the uncoated side of the third bead in the chain, which shows oscillatory behavior with time (Fig. 5c). We note that the bright parts of the beads in the chain indicate the uncoated side of the bead in the focal plane, while the dark ones indicate beads that are either above or below the focal plane. Next, we analyzed the motion of the chain in the orthogonal XZ view, which reveals the spiral motion of the chain (Fig. 5b). Fig. 5d demonstrates a typical trajectory of the center of mass of the active semi-flexible bead chain during a time interval of 20 seconds. The resulting conformations and propulsion motion is a consequence of a coupling between the propulsion forces, chain flexibility and viscous drag from the solvent that generates controlled deformations. As our chains are semiflexible in nature (persistence length, $l_p \approx 20\ \sigma$; contour length $l_c \approx 6\ \sigma$), we expect that there is always some correlation between the neighboring particles, since a combination of flexibility of the chain and hydrodynamic forces may generate an asymmetry in the self-propulsion direction of the beads. As a result, the rotation cycle is non-reciprocal, leading to (see supplementary Movie S2) a net translational displacement of the chain.[28,53] Next, we examined the flows generated in the fluid by a self-propelled 6-bead chain, by means of a 1.0 *μm* tracer particles and a 5.0 *μm* elongated glass piece (a contamination in the sample cell). The typical trajectories of tracers over 20 sec are shown in Fig. 5e. One clearly deduces from the paths of the tracer particles that there is no bulk convective flow present in the system.

In the presence of 1.0 *vol%* $H_2O_2$, the chain showed both rotational and translational motion. The active rotational velocity around the axis of the active 6-bead semiflexible chain (Fig. 5) is 3.5 rad/sec and the chain is propelling along the long axis with a velocity of 0.9 μm/s. The chain of 8 beads propelled with a translational velocity of 0.62 μm/s and the rotational velocity around the axis is 3.8 rad/sec in 1.0 *vol%* $H_2O_2$ solution. Due to the large persistence length ($l_p \approx 20\ \sigma$) of the semiflexible chains, the shorter chains ($\leq 4\sigma$) behaved like rigid chains and showed rotator behaviour.

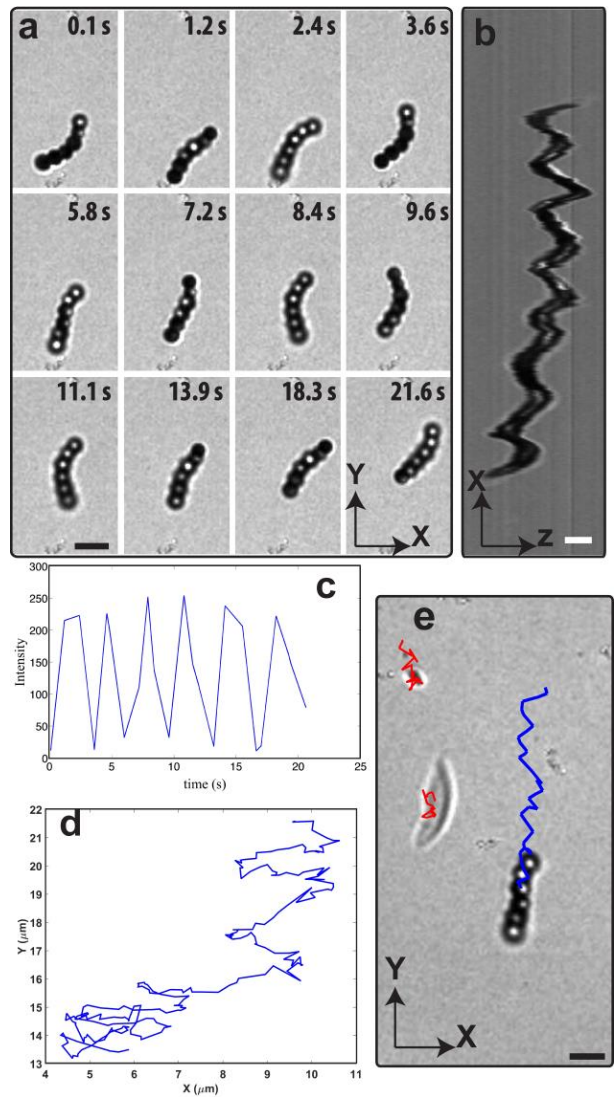

**Figure 5**. Dynamics of a self-propelled semiflexible colloidal bead chain in a 1.0 vol% $H_2O_2$ solution. (a) Time-lapse optical microscopy images show the sequential shape and orientation changes of the active bead chain resulting in net translational motion. We note that the bright parts of the beads in the chain represent uncoated sides of the beads in the focal plane while the darks ones represent beads either above or below the focal plane. (b) Orthogonal XZ view of the same chain over 100 seconds. (c) Intensity profile of the center of mass of the chain. (d) Trajectory of the centre of mass of the active semi-flexible bead chain over a time interval of 20 sec. (e) Blue line show the trajectory of the center of mass of semiflexible chain, and red lines show the trajectory of the tracer particle of a 1.0 μm size tracer particle and a 5.0 μm elongated glass piece (a contamination in the sample cell), respectively, overlaid on a bright field image. Scale bar are 5.0 μm in (a,e) figures, 2.0 μm in figure (b).

In addition, we will demonstrate now that our method can also be used to make more complex self-propelled chains composed of both semiflexible and rigid parts as shown in Fig. 6. These complex chains were fabricated by mixing rigid and semiflexible chains together and repeating the same protocol as was used in making the constituent bead chains. In the presence of fuel (1.0 *vol%* $H_2O_2$), the propelling behaviour (see supplementary Movie S3) of a half-rigid and half-semiflexible 8-bead chain resembles the swimming behaviour of microorganisms with a flexible tail and a rigid head, such as e.g., E. coli. The rotational velocity of the long axis of the half-rigid and half-semiflexible 8-bead chain is 1.9 rad/sec and the propelling velocity is 1.2 µm/s. We believe that our method opens up new possibilities to realize more complex swimmers, for instance, attaching a semiflexible chain to a 'passive' single big sphere or a cargo.

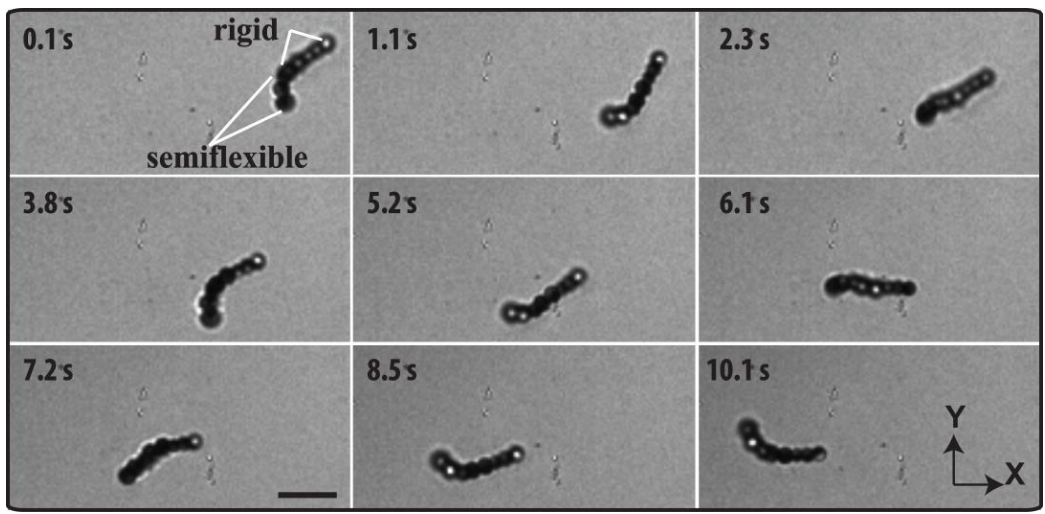

**Figure 6**. Dynamics of a self-propelled complex colloidal bead chain consisting of a half-rigid and a half-semiflexible part of the chain in a 1.0 vol% $H_2O_2$ solution. Time-lapse optical microscopy images show the swimming behaviour. White lines depict the rigid and the semiflexible parts of the chain. Scale bar is 5.0 µm.

**Bead-shell minimal model and equations of motion**

In order to better understand the experimentally observed propulsion behavior of bead chains, we complemented our study with a minimal model that includes full hydrodynamic interactions with the fluid (see more details in Materials and Methods section). Instead of modeling the catalytically powered self-propulsion in terms of a slip velocity on the surface of the Janus particles, we model the propulsion mechanism by an effective propulsion force of a fixed magnitude that acts on each of the Janus spheres. (see more details in Materials and Methods section). For a rigid chain of spheres, these propulsion forces sum up to a force and torque acting on this rigid configuration. If the $6 \times 6$ hydrodynamic resistance tensor of this rigid body is known, the equations of motion of this object can be solved after imposing the effective force and torque. Alternatively, if the motion of the body is known, these equations of motions may be inverted to calculate the effective force and torque. We apply this to the rigid rotating chains, for which the observed angular velocity is used to calculate, given the 'zig-zag' arrangement of the propulsion forces, the magnitude of the effective propulsion force (more details in Materials and Methods section). To check the validity of this procedure, we feed the obtained effective force magnitude back into the equation of motion, from which we calculate the angular velocity for rigid bead chains of different lengths. We found our calculations to be consistent with the experimentally measured angular velocities (Fig. 3) in all three cases of bead chains with an even number of beads, as shown in Fig. 7.

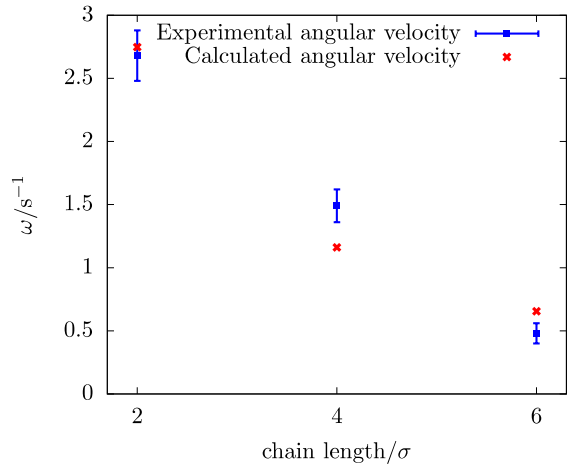

**Figure 7**. The experimentally measured angular velocity for bead chains of different length in comparison with the theoretical ones obtained from the average effective propulsion force using the resistance tensor.

**Self-propelled semi-flexible bead chains**

As shown experimentally, the active rigid bead chains did not exhibit any translational motion (Figs. 1b and 2), whereas the semiflexible chains showed a net translational motion (Fig. 5). Due to the over-damped nature of the dynamics at low Reynolds numbers, we expect that the internal motions within a semi-flexible bead chain will quickly decay towards an equilibrium configuration, rather than showing oscillatory motion as in the case of a driven oscillator with inertia. Below we show that a small anisotropy in both the shape and the force distribution of the bead chain, induced by the semiflexibility, can lead to a translational motion of the bead chain, even when the net effective propulsion force on the chain remains zero. To this end, we analyzed small anisotropies in the shape by considering bead chains with the centers of the Janus spheres positioned in a C-shape and on a helical centerline, as parameterized by a helical radius $r$ and helical pitch $p$ (see more details in Materials and Methods section). In addition, we studied anisotropic arrangements of these propulsion forces that deviate from the 'zig-zag' arrangement.

We summarized our numerical predictions in terms of 'swimming' behavior for the active helical bead chains with varying shape and propulsion force distributions in the Figure-T I. We clearly found that anisotropy in both shape and force distribution is required to induce translational motion.

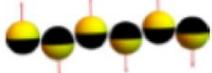

**Figure-T I.** Summary of whether or not translational motion is predicted theoretically, as indicated by Yes and No, for active helical bead chains with varying shapes and force distributions. We distinguish three different shapes (the `zig-zag' linear bead chain, the C-shaped chain, and different types of helically shaped bead chains) and two different arrangements of the effective propulsion forces (an alternating `zig-zag' force distribution as observed for the rigid bead chains and a typical heterogeneous force distribution). Animations of the motion of these different bead chains, referred to in parentheses, can be found in the
Supplementary Movies.

Hence, shape anisotropy in the form of a C-shaped or (moderately) helically shaped bead chain, combined with a heterogeneous distribution of propulsion forces, leads to a 'swimming' motion that qualitatively reproduces the experimentally observed propulsion behavior (Fig. 5).

This can also be clearly seen in animation 6 found in the Supplementary information. We now take a close look at the influence of the force distribution on the swimming behavior. In Fig. 8, we compare the in-plane (projected) velocity $v_{2D}$ and angular velocity around the long axis of the bead chain, $\omega_{\parallel}$, for a set of 100 randomly chosen configurations of propulsion forces for a fixed helically shaped bead chain, to the (angular) velocity for a range of different helical bead chains and the experimentally observed values. Here, we considered configurations with 6 randomly chosen orientations (blue), which are (almost) never force-free. In addition, we considered zig-zag configurations with a single randomly chosen orientation (purple), as well as configurations with three randomly chosen directions of propulsion forces, augmented with the three opposite vectors. For the latter two categories, the total effective propulsion force vanishes. From Fig. 8, we observed that these random configurations do not produce translational motion that qualitatively agrees with the observed experimental motion and therefore we conclude that the arrangement of the effective propulsion forces requires a certain correlation of the directions of the forces on the beads. Since the experimental system is semiflexible in nature (persistence length, $l_p \approx 20\sigma$; contour length $l_c \approx 6\sigma$) and initially made from a linear zig-zag beadchain, there are always correlations between the force directions of neighboring beads in the chain.

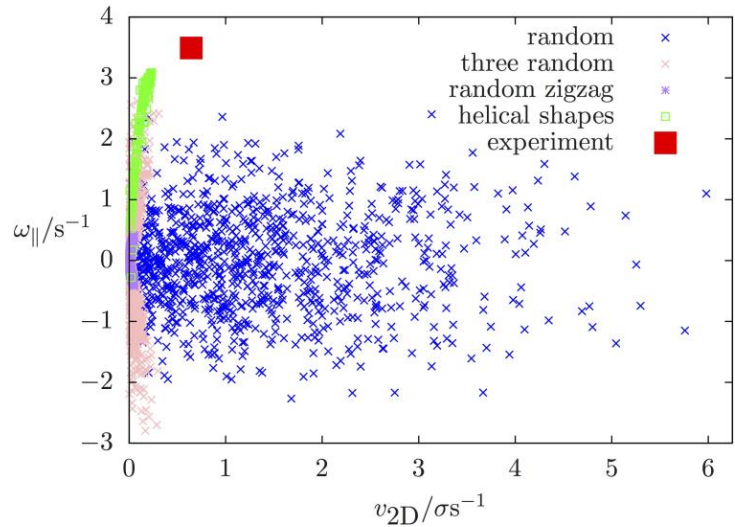

**Figure 8**. Comparison of the (magnitude of the) two-dimensional velocity $v_{2D}$ and angular velocity $\omega_{\parallel}$ for different distributions of the propulsion forces on the beads of a helical bead chain. In blue, 100 realizations are shown where the direction of the propulsion force is chosen at random for each individual Janus particle. Force-free (i.e. vanishing effective propulsion force) realizations with a `zig-zag' distribution with forces aligned anti-parallel in a random direction, and where the direction of the propulsion forces is chosen at random for three of the Janus particles with the three others being their opposite, are shown in purple and pink, respectively. These numbers were calculated for a specific choice of helical bead chain shape, as shown in Fig. 9. In green, the (angular) velocity is plotted for the above mentioned heterogeneous force distribution, for all the different helically shaped bead chains under consideration, while the red square indicates the values that are observed in the experiment.

**Conclusions**

In conclusion, we have developed an effective, yet simple method for the fabrication of internally powered colloidal bead chains with tunable stiffness. Our active bead chains are the first experimental realization of a new type of self-propelled particles that spontaneously rotate or spin when the spherical swimmers are rigidly connected, while they show flagellum-like propulsion when the connections between the spherical swimmers are semiflexible. In the presence of the fuel, active rigid bead chains experience equal and opposite propulsion forces along the length of the chain that induce active rotational motion. Due to the hydrodynamic forces and the flexibility of the chain, long ($\geq 6\sigma$) self-propelled semiflexible chains are subject to asymmetry in the self-propulsion direction of the beads in a chain, leading to flagellum-like motion. We corroborated our experimental results by means of numerical calculations with a minimal theoretical model, based on a description via effective propulsion forces. We were able to reproduce qualitatively and to some extent quantitatively, the 'swimming' behavior of the semiflexible bead chains. Also, the spiraling motion calculated from this model agrees well with the experimentally observed motion of the bead chains. With our numerical calculations we found that the propulsion forces along the chain requires some correlations between the directions of the forces on the beads to induce translational motion for active semiflexible chains. These correlations may arise from the semiflexible nature of the chains that initially had a well-defined linear zig-zag structure. Our findings demonstrate how microscopic dynamics (i.e., flexibility between the beads) can affect the dynamical behavior of self-propelled bead chains. In future studies, the length of the chains can be controlled using previously reported methods by Vutukuri et al.[34] Potentially, our method paves a way to study collective behavior of self-propelled or internally driven chains with tunable stiffness, and can likely be extended to light-activated propelling particles[7] (e.g., half-titania coated particles) where the rotational speed of rigid chains and the swimming speed of semiflexible bead chains can be controlled with light intensity. Additionally, we can easily extend the procedure to magnetodielectric particles such as magnetic nanoparticles embedded in micron-sized polymeric particles, so that we can control the directionality of the motion of active bead chains by either external magnetic or electric fields. This work will facilitate future designs of new complex self-propelled swimmers, for instance, attaching a 'passive' single big sphere or a cargo with an active semiflexible bead chain.

**Materials and Methods**

**Particle synthesis.** We synthesized[54] electrostatically stabilized and negatively charged polystyrene (PS) particles of 1.35 µm with a size polydispersity of 4 %, and sterically stabilized (polyvinylpyrrolidone, PVP, $M_w$ = 40 kg/mol) with a size polydispersity of 3 %. These two types of particles were used in the fabrication of rigid bead chains. Whereas a longer molecular weight PVP ($M_w$ = 360 kg/mol) PS particles were used in the fabrication of semiflexible bead chains. Sterically stabilized PS particles were synthesized using the method of Song *et al.*[55] The size of the sterically stabilized PS particles was 1.40 µm with a size polydispersity of 5 %. The particle size and polydispersity were determined using static light scattering, and scanning electron microscopy. We note that based on our camera frame rate, we deemed it easier to follow the dynamics of tracer particles in the case of rigid bead chains when we used bigger particles (2.1µm), because the chains are rotating faster. On the other hand, smaller particles (1.0 µm) are sufficient to capture the dynamics of tracer particles in the case of semiflexible case.

**Janus particles preparation.** We first prepared a monolayer of spherical particles by slowly drying particles from a dilute suspension (0.05 *wt%*) on a clean microscope glass slides. A 15 nm thick layer of platinum (Pt) was then vertically deposited using a sputter coater (Cressington 208HR). Prior to particle detachment, slides were thoroughly washed with DI water, and subsequently particles were detached by sonicating the slides in DI water for 10 mins. Next, particles were washed a couple of times with DI water and then the suspension with the desired concentration was transferred to the electric cell.

**Electric-field setup**. The electric cell consisted of a 0.1 mm x 1.0 mm cross section capillary with two 50 µm thickness nickel-alloy wires (Goodfellow) threaded along the side walls[56]. We used a function generator (Agilent, Model 3312 OA) and a wide band voltage amplifier (Krohn-Hite, Model 7602M) to generate the electric fields. We used a high frequency AC fields to minimize the polarization effects of the electric double layer around the particle. The field strength and the frequency were measured with an oscilloscope (Tektronix, Model TDS3052). We used a homemade DC filter to remove the DC component in the signal.

**Fixation of bead chains**. After filling the cell with the Janus colloidal particles suspension, we sealed both ends of the capillary with UV-curing optical adhesive (Norland No. 68) and cured the glue with UV light ($\lambda$ = 350 nm, UVGL-58 UV lamp). The dispersion was then exposed to an AC field ($E_{rms}$ = 0.04-0.05 V$\mu$m$^{-1}$, $f$ = 800 kHz), and the ramping time was about 2 mins. After 2-3 mins, all the particles were assembled into zig-zag chains of one particle wide in the field direction. The dispersion was subsequently heated to 60-65 °C, which is well below the glass transition temperature ($T_g \approx$ 107 °C) of polystyrene,[47] for about 2-3 minutes using a hot-air stream that was much wider than the sample cell.[34,39,57] We further kept the field on for 2-3 mins while the sample was cooled down to room temperature. After the fixation step, we carefully opened both ends of the electric cell and the bead chains were subsequently transferred to a small eppendrof tube (0.5 ml) for further dilution. The dynamics of the chains were studied in Lab-Tek chambered cover-glass (Thermo Fisher Scientific).

**Internal structure of staggered or zig-zag chains**. Our method yielded regular and staggered (zig-zag) linear chains independent of whether it is used in the fabrication of rigid or semiflexible chains. We note that in our fabrication method, we exploited the difference in the polarizability of both sides (half-dielectric and half-metallic) of the Janus particles in external AC electric fields, causing a zig-zag chain configuration, as is shown in the SFig.1. In the case of semiflexible (persistence length, $l_p \approx$ 20 $\sigma$; contour length $l_c \approx$ 6 $\sigma$), we expect that there remains always some correlation between the neighboring particles, stemming from the zig-zagged pattern at synthesis, while a combination of flexibility of the chain and hydrodynamic propulsion may generate an asymmetry in the self-propulsion direction of the beads.

**Particle tracking.** We recorded the particle dynamics using an Olympus IX81 confocal microscope, equipped with an Andor iXon3 camera, Andor 400-series solid-state lasers, a Yokogawa CSU-X1 spinning disk. Bright-field optical micrographs and videos were recorded using a Nikon microscope equipped with a CCD camera (Olympus CKX41). We processed recorded images and extracted the centroids of each particle in a chain using the particle tracking programs of Rogers *et al.*,[50]. We then calculated the orientation of the chain from the pixel coordinates of the first bead and the last bead in the chain. To uniquely identify the both ends of the chains, it is required that the rotation angle of the chain between successive frames is less than $\pi/2$.

**Mean squared displacement of rigid chains.** The MSAD and MSDs can be determined by solving the equations of motion for a circular swimmer.[52] The overdamped equations can be written in 2D as

$$\frac{dx(t)}{dt} = v \cos\theta(t) + \xi_1(t) \qquad (1)$$

$$\frac{dy(t)}{dt} = v \sin\theta(t) + \xi_2(t) \qquad (2)$$

$$\frac{d\theta(t)}{dt} = \omega + \zeta(t) \qquad (3)$$

where $\omega$ is the rotational velocity, $x$ and $y$ are the center of mass coordinates, $v$ is the propulsion velocity, and $\theta$ is the orientation of the bead chain. The Brownian noise terms, $\xi$, $\zeta$, are Gaussian random variables with zero mean whose magnitudes are taken from theoretical Brownian diffusivities. By solving the Eqs. 1-3 we obtained

MSAD: $\langle \Delta\theta^2 \rangle = \langle [\theta(t) - \theta(0)]^2 \rangle = \omega^2 t^2 + 2 D_r t$ (4)

MSD: $\langle \Delta L^2 \rangle = \frac{2v^2}{(D_r^2 + \omega^2)^2} [(D_r^2 + \omega^2) D_r t + (\omega^2 - D_r^2) + e^{-D_r t}([D_r^2 - \omega^2] \cos\omega t - 2\omega D_r \sin\omega t)] + 2(D_\parallel + D_\perp) t$ (5)

where $D_r$ is the rotational diffusion coefficient, $D_\parallel$, $D_\perp$, are translational diffusion coefficients along the major and minor axis, and $v$ is the propulsion velocity.

**Resistance tensor, bead-shell model and equations of motion.**

Often, the catalytically powered self-propulsion of (half-)platinum coated particles, such as the bead chains considered this work, is modelled by a slip-velocity on the active part of the particle surface. Moreover, since there are no external forces acting on the bead chains, the total hydrodynamic force and torque on the chain both vanish, such that the flow field around the bead chain is, to leading order, at most of dipolar character. However, as was proven by Ten Hagen et al. [58], the *motion* of a self-propelled particle is equivalent to that of a passive particle (of identical shape), which is driven by an *effective* external force ***F**^{eff}* and torque ***T**^{eff}*. As described in Ref 58, if both the distribution of slip-velocity on the active particle surface, and flow field solutions around the driven passive particle for arbitrary external force and torque are known, the motion of the active particle may be solved by taking suitable choices of the driving forces in the conjugate passive particle system. Subsequently, the effective force $\boldsymbol{F}^{eff}$ and torque $\boldsymbol{T}^{eff}$ are calculated from the passive case by imposing the velocity and angular obtained in the previous step. Due to the linearity of low-Reynolds number hydrodynamics, the force and torque relate to the particle velocity $\boldsymbol{U}$ and angular velocity $\boldsymbol{\omega}$ in a linear fashion:

$$\mathcal{F} = \eta\, \mathcal{R}\, V \qquad (6)$$

where the 6-vector $\mathcal{F} = (\boldsymbol{F}^{eff}, \boldsymbol{T}^{eff})$ denotes the (effective) force and torque on the particle, the 6-vector $V = (\boldsymbol{U}, \boldsymbol{\omega})$ denotes the (angular) velocity with respect to a chosen reference frame and the 6 x 6 tensor $\mathcal{R}$ denotes the hydrodynamic resistance tensor, which depends itself on the shape of the self-propelled particle under consideration. The linearity of this dependence is due to the linearity of the Stokes equation that governs the hydrodynamics at low Reynolds numbers.

Here, rather than calculating the effective force and torque from the slip-velocity on the active bead chains, and flow field solutions of the passive problem, we adopt the opposite strategy by directly determining the effective force and torque from bead chain systems for which both the motion and the geometry are accurately known (from the experiment). To this end, we numerically (using a bead-shell model, see p. 21) determine the tensor $\mathcal{R}$ for the rigid bead chains, for which the (angular) velocity is measured in the experiment: $\boldsymbol{U} = \boldsymbol{0}$, while $\boldsymbol{\omega} = \omega\hat{\boldsymbol{z}}$, which leads to $\boldsymbol{F}^{eff} = 0$ and $\boldsymbol{T}^{eff} = T^{eff}\hat{\boldsymbol{z}}$, where $T^{eff}$ is solved from the equation of motion $\mathcal{F} = \eta\, \mathcal{R}\, V$, after we determined the resistance tensor $\mathcal{R}$.

To proceed, we decompose the effective force and torque into effective forces acting on each of the individual Janus sphere that constitute the chain, as

$$\boldsymbol{F}^{eff} = \sum_i \boldsymbol{F}_i \quad \text{and} \quad \boldsymbol{T}^{eff} = \sum_i \boldsymbol{F}_i \times (\boldsymbol{r}_i - \boldsymbol{r}_{cm}) \qquad (7)$$

where we assume that the force magnitude $|\boldsymbol{F}_i|$ is equal for all $i$, while the direction is determined by the orientation of the catalytic hemispheres: pointing along the axis that connects the `south' pole on the passive hemisphere to the `north pole' of the active hemisphere, as is shown in e.g. Fig. 9 of the main text. This assumption is equivalent to assuming that the (chemical) interaction between catalytic surfaces is small, such that all $\boldsymbol{F}_i$ are of equal magnitude, and that the symmetry of the active hemisphere is preserved such that (effective) contributions of individual torques on each Janus sphere may be neglected, as would be the case of a single Janus sphere. Since an alternating (`zig-zag') orientation of the active hemispheres is clearly observed in the experiment, the magnitude $|\boldsymbol{F}_i|$ is the only unknown and may be solved by plugging Eq. (7) into Eq. (6) for the rigid rotating chains, as described in the next section.

In this work, we calculate the rigid body resistance tensor of (active) bead chains using a bead-shell model,[59,60] in which the surface of a rigid body is homogeneously covered by a large number *(M)* of small spheres of radius *a*. When this cluster of *M* spheres is given a non-zero common velocity, a disturbance flow field is created, which in turn causes hydrodynamic interactions between the small spheres that are given by the Rotne-Prager mobility tensor [61,62]. The forces on the individual spheres can then be calculated by a 3*M* x 3*M* matrix inversion, from which the total force and torque on the rigid body follow as the sum of the individual forces and torques around a chosen reference point. Subsequently, these results are extrapolated to $M \to \infty$, while keeping $Ma^2$ finite, where typically *M* =1000-3000 spheres are used in this procedure. In this large-*M* limit, we retrieve the boundary integral formulation of the Stokes equation, guaranteeing accurate results for the resistance tensor $\mathcal{R}$.

**Self-propelled rigid bead chains.** We first validate our model for the active rigid bead chains by comparing the rotational diffusion coefficient for the 6-bead chain as obtained from the bead-shell model calculations to the experimental data. Assuming that the direction perpendicular to the plane of view of the microscope is the *z*-axis, the relevant rotational resistance coefficient is $\boldsymbol{R}_{66}$, from which the rotational diffusion coefficient is expressed by the Stokes-Einstein relation as $D_r = k_B T / (\eta\, R_{66})$. Here, using the viscosity of water and the experimental size of the Janus spheres, we find $D_r = 0.0096$ rad$^2$/s, which is consistent with the experimentally measured value. We included the experimental results of rotational speed of the rigid chains in the presence of the fuel and the rotational diffusivity of chains in water, and the rotational and the translational speed of semiflexible chains in the fuel in supplementary information (see Table SII).

For completeness, we give the complete resistance tensor in the below paragraph. Next, we use the measurements of the rotational velocity of the rigid bead chains to estimate the value of the effective propulsion force acting on each individual Janus sphere. Note that in each case, we try to stay as closely as possible to the observed 'zig-zag' shape, as shown in SFig. 5, where we estimate from experimental snapshots that the angle between the vectors connecting the centers of adjacent Janus spheres is $0.15\pi$ on average. Also, the distribution of the propulsion forces is arranged in an alternating zig-zag fashion.

The validity of our model and consistency of our assumptions, Eq. (7), is verified from Fig. 7

above, where we compare the experimentally observed angular velocity $\boldsymbol{\omega}$ with the angular velocity that we calculate from Eq. (6) for chains of different length, where the values of $|\boldsymbol{F}_i|$ is averaged over the data of the rigid bead chains.

The complete 6x6 resistance tensor of a rigid linear 'zigzag' bead chain of 6 beads, aligned along the $x$-axis, is found to be:

$$\mathcal{R} = \begin{pmatrix} 38.0\ \mu m & -0.35\ \mu m & -0.042\ \mu m & 0.097\ \mu m^2 & -0.0041\ \mu m^2 & -0.015\ \mu m^2 \\ -0.35\ \mu m & 30.0\ \mu m & -0.056\ \mu m & -0.005\ \mu m^2 & -0.01\ \mu m^2 & 0.022\ \mu m^2 \\ -0.042\ \mu m & -0.056\ \mu m & 38.0\ \mu m & 0.022\ \mu m^2 & -0.0039\ \mu m^2 & -0.09\ \mu m^2 \\ 0.097\ \mu m^2 & -0.005\ \mu m^2 & 0.022\ \mu m^2 & 430.0\ \mu m^3 & -11.0\ \mu m^3 & 1.1\ \mu m^3 \\ -0.0041\ \mu m^2 & -0.01\ \mu m^2 & -0.0039\ \mu m^2 & -11.0\ \mu m^3 & 46.0\ \mu m^3 & 0.15\ \mu m^3 \\ -0.015\ \mu m^2 & 0.022\ \mu m^2 & -0.09\ \mu m^2 & 1.1\ \mu m^3 & 0.15\ \mu m^3 & 430.0\ \mu m^3 \end{pmatrix}$$

**Self-propelled semi-flexible bead chains.** With the magnitude $|\boldsymbol{F}_i|$ determined from the rigid chains and the consistency of our assumptions checked, we apply this model to the motion of the semiflexible bead chains. As mentioned above, we model the semiflexible bead chains by introducing shape anisotropy to the rigid bead chains, such that the centers of the beads lie on a helix with radius r and pitch p. We note that the aligned chain ($p \rightarrow \infty$) and the c-shaped chain ($p = 0$) are special cases of this parameterization.

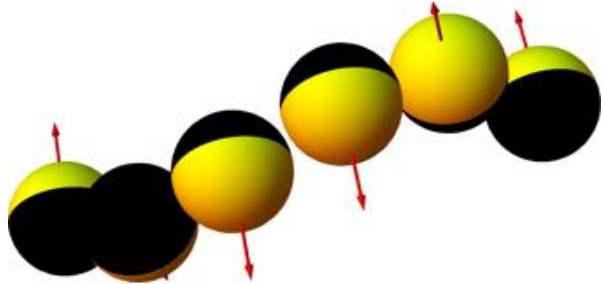

**Figure 9**. A (constructed) configuration of effective propulsion forces that gives rise to a rotation of the bead chain. The forces are all perpendicular to the axis that connects the two end beads, which is lying on a horizontal plane in the experiment. The forces differ by a rotation around this axis, such that the resulting rotation is predominantly around this axis.

Moreover, we consider a propulsion force distribution along the bead chain that is arranged in such a way that the chain will start rotating around the axis that connects the two ends, when the chain does not coincide completely with this axis. The propulsion forces are all perpendicular to the axis that connects the end points of the bead chain, and differ by a rotation around this axis. Below, we denote the angular velocity around this axis by $\omega_{//}$, while we represent the projected (2-dimensional) velocity in the xy-plane by $v_{2D}$. Note that, as in the case of the rotating rigid bead chains, the net force on the chain vanishes. Therefore, any translational motion is entirely caused by a (shape-dependent) translation-rotation coupling in the resistance tensor described above, and is in fact independent of our effective description using propulsion forces (that we have obtained from the rigid case).

We make a comparison between the theoretically calculated velocities associated with these bead chains and the velocity measured in the experiment, which is $0.9\,\mu m/s = 0.64\,\sigma/s$. In SFig.6, we plot the velocity of helical bead chains as a function of the helical radius $r/\sigma$ and pitch $\alpha$ as obtained theoretically using our bead-shell model. We observe that over a large range of shapes, the velocity is around $0.2\,\sigma/s$. We also investigate the rotational velocity $\omega_{//}$ for the same range of shapes, as shown in SFig.7. Here, we observe a wide range of shapes that show an angular velocity $\omega_{//}$ of $3.0\,s^{-1}$, which compares well to the value of $3.5\,s^{-1}$ in the experiments.

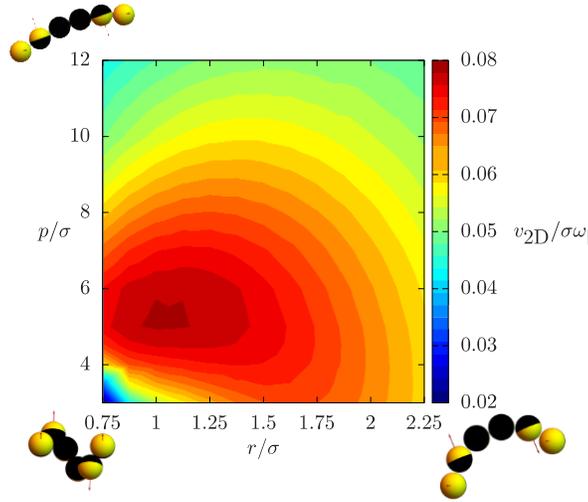

**Figure 10**. Ratio between the in-plane velocity $\mathbf{v}_{2D}$ and the angular velocity $\omega_{//}$ around the bead chain axis, for a range of helically shaped bead chains characterized by the helical radius r/σ and pitch p/σ.

Finally, we studied the ratio between rotation and translation velocity, which as described above, is purely an effect of the shape of the bead chain and is independent of the magnitude of the effective propulsion force. In Fig.10 we show this ratio as a function of helical shape, and observe that this ratio is around $0.08\,\sigma$ for a range of helical shapes, whereas the experimentally measured ratio is $0.18\,\sigma$. We conclude that our minimal model is able to predict this ratio up to a factor of two, which is striking given the simplicity of the model. Also, the spiraling motion observed from the animations coincides well with the experimentally observed motion. A better understanding of experimental uncertainties or a more detailed hydrodynamic model may improve these predictions.

**Acknowledgements**

We would like to thank Frank Smallenburg for useful discussions. W.T.S.H. acknowledges financial support by the European Research Council Advanced Grant (246812 intercom), H.R.V. was partially supported by a Marie Skłodowska-Curie Intra European Fellowship (G.A. No. 708349- SPCOLPS) within Horizon 2020, and R.v.R. and B.B. by a NWO-VICI grant of the Netherlands Organization for Scientific Research.

**Author contributions**

H. R.V. conceived the research and designed the experiments. H.R.V. performed all the experiments and analysis. W.T.S.H. supervised the research. B.B., R.v.R., and M.D. developed the numerical models. B.B. performed the calculations. H.R.V., B.B., and W.T.S.H. wrote the manuscript. All authors discussed the results and commented on the manuscript.

**Additional information**

**Supplementary Information** accompanies this paper at http://www.nature.com/naturecommunications

**Competing financial interests.** The authors declare no competing financial interests.